# Semiconductor nanowires studied by photocurrent spectroscopy

**The copyrights of the text are with Woodhead publishing. It will be part of the book, entitled**

**"Semiconductor nanowires materials, synthesis, characterization and applications"**

See: www.woodheadpublishingonline.com

**Authors:** Nadine Erhard[1,2] and Alexander Holleitner[1,2]

1 Walter Schottky Institut and Physik-Department, Technische Universität München, 85748 München, Germany
2 Nanosystems Initiative Munich (NIM), Schellingstr. 4, 80799 München, Germany

## Abstract

Photocurrent spectroscopy is a versatile technique to identify and understand the optoelectronic dynamics occurring in semiconductor nanowires. Conventional photocurrent spectroscopy allows to explore the morphology and material properties of nanowires as well as their contact interfaces. Using time-resolved photocurrent spectroscopy one gets additional information on the multiple photocurrent generation mechanisms and their respective timescales. This chapter discusses various aspects of the photocurrent spectroscopy and it summarizes the physical mechanisms behind the photocurrent and photoconductance effects in semiconductor nanowires.

# 1. Introduction

First investigations of photocurrents in semiconductor nanowires were performed at the beginning of the millennium (Duan et al. 2001; Hayden et al. 2006; Keem et al. 2004; Kind et al. 2002). Since then nanowires have been increasingly investigated with respect to future applications in solar cells (Kempa et al. 2008; Tian et al. 2007), photodetectors (Bulgarini et al. 2012; Hayden et al. 2006; Prechtel et al. 2012a) and optical sensors (Falk et al. 2009; Lee et al. 2004; Lee et al. 2007). Photocurrent spectroscopy is a powerful tool to identify the generation mechanisms of the photocurrent in nanowire-based devices (Erhard et al. 2013; Prechtel et al. 2012a). The observed photocurrent effects can be significantly influenced by the nanowire geometry. For instance, the absorption can be dominated by the dielectric confinement (Cao et al. 2009; Wang et al. 2001). Nanowires can be grown with diameters below the Bohr radius, resulting in one dimensional quantum confined structures (Law et al. 2004). Furthermore, nanowires with crystal structures not inherent to the bulk have a different band structure and band gap (Maharjan et al. 2009; Trägårdh et al. 2007). Strain free heterostructures with radial or axial symmetries allow to build new types of optoelectronic nanowire devices, inherently different to circuits based on thin film technologies (Kuykendall et al. 2007; Li et al. 2010). Due to the large surface to volume ratio photoconductance effects negligible in bulk semiconductors can further dominate the optoelectronic response of semiconductor nanowires. This chapter is structured as follows: Section 2 describes the conventional and time-resolved photocurrent spectroscopy. Section 3 summarizes the photocurrent and photoconductance effects which have been observed in semiconductor nanowires. Section 4 reviews the absorption effects in nanowires investigated by photocurrent spectroscopy. Section 5 discusses the photocurrent and photoconductance effects for the different nanowire morphologies. The chapter concludes with an outlook in section 6.

# 2. Methods

## 2.1. Conventional photocurrent spectroscopy

Conventional photocurrent spectroscopy explores the photocurrent of electrically contacted semiconductor nanowires as a function of e.g. excitation wavelength, bias voltage, and light polarization. To this end, a laser is focused through an objective onto the nanowire-based circuit and a bias voltage is applied to one of the contacts. In some experiments, a chopper is used to enhance the signal to noise ratio (Dunn 2000) and to get access to optoelectronic processes in the frequency range of Hz to kHz (s to ms).

Up to now, optical absorption experiments on a single nanowire are still challenging and have only been realized for high illumination intensities (Gabriel et al. 2013). Photocurrent spectroscopy, however, enables absorption measurements on single nanowires also for low intensities (Chen et al. 2013),  albeit the small fraction of light which is absorbed by a single nanowire (Cao et al. 2009; Kim et al. 2013). Hereby, one can get information on the nanowires' band structure and joint density of states (Bennett 1971). Thus, photocurrent spectroscopy can be a good supplement to photoluminescence measurements, as photocurrent spectroscopy can be easily performed at room temperature and it is also applicable to indirect band gap materials (Chen et al. 2013).

## 2.2. Time-resolved photocurrent spectroscopy

Conventional photocurrent studies are typically limited to timescales exceeding 10 ps (Gallo et al. 2011) because available electronic equipment cannot produce and detect faster trigger signals and transients. Furthermore, optoelectronic charge-carrier dynamics are obscured by the response time of the high-frequency circuits. Yet, it is known from optical experiments that carrier relaxation, thermalization, and recombination processes can occur on much faster time scales in semiconductor nanowires (Johnson et al. 2004). Using time-resolved photocurrent spectroscopy one can investigate the dynamics of photo-thermoelectric currents, displacement currents, the transport of photogenerated charge carriers to the electrical contacts, and the photo-Dember effect, as well as the carrier lifetime limited currents with a picosecond time-resolution (Figure 1). In time-resolved measurements, the individual current contributions can be separated because of their characteristic timescales and dynamics in nanowires. Moreover, the drift velocity of the photogenerated holes can be explored (Erhard et al. 2013; Prechtel et al. 2012a). The results obtained by time-resolved photocurrent spectroscopy prove useful for the design of nanowire based photodetectors, photoswitches, solar cells, and high-speed transistors.

There exist two major pump-probe methods to measure the time-resolved photocurrent on a picosecond timescale. These are the photocurrent autocorrelation spectroscopy (Erhard et al. 2013; Gabor et al. 2012; Urich et al. 2011) and the THz-time domain photocurrent spectroscopy (Erhard et al. 2013; Prechtel et al. 2012a).

### 2.2.1.  Photocurrent autocorrelation spectroscopy

For the photocurrent autocorrelation spectroscopy, a pump laser pulse is focused on the nanowire (Figure 2). This exciting pulse needs to be intense enough to generate a photocurrent in saturation (e.g. because of bleaching due to state filling (Ahn et al. 2012; Li et al. 2014; Lo et al. 2012)). A second probe laser pulse hits the nanowire at the same position after a time delay $\Delta t$. The probe laser pulse can only excite additional charge carriers and hereby an additional photocurrent if the saturation photocurrent induced by the pump laser pulse is already partially relaxed after $\Delta t$. Hereby, photocurrent relaxation times in the picosecond regime have been explored (Erhard et al. 2013). However, photocurrent effects which do not saturate such as the photo-thermoelectric effect (section 3.2) cannot be explored by the photocurrent autocorrelation spectroscopy.

### 2.2.2. THz time-domain photocurrent spectroscopy

Another possibility to measure the photocurrent of a nanowire with a picosecond time-resolution is the so-called THz time-domain photocurrent spectroscopy (Erhard et al. 2013; Prechtel et al. 2012a). Compared to the photocurrent autocorrelation spectroscopy, it also works in the non-saturation regime, because the time-resolution is achieved by the utilization of an Auston switch (Figure 3). First, the electrically contacted nanowire is excited by a femtosecond laser pump pulse. The photogenerated currents in the nanowire induce an electromagnetic transient in the strip lines, by which the nanowire is contacted (optical microscope image in Figure 3). The transient is directly proportional to the photocurrent and it propagates along the strip lines to the field probe where the so-called Auston-switch is excited by a probe laser pulse at a time delay $\Delta t$. The Auston-switch is a silicon photo-switch with a sub-picosecond response time (Auston 1983; Doany et al. 1987). If at the time delay the electromagnetic transient is present at the Auston-switch, the photogenerated charge carriers in the silicon amount to a current $I_{\mathrm{sampling}}$ in the field probe. Hereby, the photocurrent of the nanowire can be sampled as a function of $\Delta t$ with a picosecond time-resolution (Erhard et al. 2013; Prechtel et al. 2012a). The THz-time-domain photocurrent spectroscopy allows to characterize nanowires in an all-encompassing way as the conventional photocurrent spectroscopy but with a time-resolution of picoseconds (Prechtel et al. 2011; Prechtel et al. 2012b). In particular, the THz time-domain spectroscopy can be used to explore photo-thermoelectric effects, displacement and drift currents (photovoltaic effects), recombination processes and THz oscillations in single nanowires (Erhard et al. 2013; Prechtel et al. 2012a).

# 3. Photocurrent dynamics in semiconductor nanowires

In principle, photocurrents involve two main physical processes: the optical absorption and the charge transport (Chen et al. 2011; Chen et al. 2013; Sze et al. 2007). However, one can distinguish several optoelectronic dynamics with different timescales (Figure 1) as will be discussed in the following sections.

## 3.1. Photovoltaic effects

In a naïve picture, the photovoltaic effect is the generation of a voltage when a device is exposed to light (Sze et al. 2007). To achieve this in a nanowire-based device, an intrinsic electric field e.g. due to a space charge region has to be present. In turn, a photocurrent is generated along the forward direction of the space charge region (Kittel 1995). In literature, the photovoltaic effect in nanowires has been investigated for a broad variety of cases as will be discussed in the following.

At a metal-semiconductor interface, a space charge region may occur (Sze et al. 2007). By illuminating this Schottky contact, a photocurrent is generated. This photovoltaic effect has been observed for many different nanowires made of CdS (Gu et al. 2005), GaAs (Thunich et al. 2009), GaN (Deb et al. 2006), Ge (Kim et al. 2010), Si (Ahn et al. 2005) and ZnO (Heo et al. 2004; Keem et al. 2004) contacted by metals, such as Au (Keem et al. 2004; Thunich et al. 2009), Ni (Ahn et al. 2005), Pt (Deb et al. 2006; Heo et al. 2004) and Ti (Gu et al. 2005). In addition, at the interface between a semiconductor nanowire and a zero band gap material, such as graphene, a photovoltaic effect can be observed and the results can be equally explained in terms of a Schottky barrier (Dufaux et al. 2010; Fan et al. 2011).

Bulk photovoltaic devices typically consist of a pn-homo-junction or hetero-junction (Sze et al. 2007). This concept was also realized in semiconductor nanowires (Garnett et al. 2011; Peng et al. 2011). Due to the nanowire geometry, however, there exist several possibilities to implement such junctions in a nanowire device (Figure 4). One way to achieve a nanowire pn-junction is to deposit a p-type nanowire on a crossing n-type nanowire (Figure 4a) (Cui 2001; Duan et al. 2001; Hayden et al. 2006). Further possibilities are photovoltaic devices made out of radial (Figure 4b) (Christesen et al. 2012; Czaban et al. 2009; Dong et al. 2009; Garnett et al. 2008; Krogstrup et al. 2013; Mariani et al. 2011; Tang et al. 2011; Tian et al. 2007; Wang et al. 2010) or axial (Figure 4c) (Christesen et al. 2012; Guo et al. 2010; Heurlin et al. 2011; Kempa et al. 2008; Li et al. 2009; Lin et al. 2011; Reimer 2011; Sivakov et al. 2009) pn-junctions within the nanowires. Moreover, p-doped (n-doped) substrates with n-doped (p-doped) nanowires grown on top can also exhibit a photovoltaic effect (Figure 4d) (Garnett et al. 2010; Peng et al. 2005; Stelzner et al. 2008; Tang et al. 2008; Wei et al. 2009). For n-type ZnO nanowires combined with p-type CdSe (Leschkies et al. 2007) or $CuO_2$ (Yuhas et al. 2009) nanoparticles, photovoltaic effects have been reported. Moreover, solar cells have also been built by embedding an n-type nanowire array in a thin film of a p-type semiconductor (Bie et al. 2010; Fan et al. 2009; Lévy-Clément et al. 2005).

Last but not least, bending a nanowire induces mechanical strain which causes internal electric fields. Hereby, photogenerated electrons and holes are seperated (Greil et al. 2014; Wu et al. 2009). Such a strain-induced photovoltaic effect was observed in Ge nanowires (Greil et al. 2014).

The photovoltaic effect can generate photocurrents on different timescales. On short timescales, in the first picoseconds after a pulsed laser excitation, the generation of additional charge carriers may cause a displacement current by which electric fields are screened. On timescales of picoseconds up to nanoseconds, the photocurrents can be dominated by the drift current of slow photogenerated charge carriers, such as holes, and the carrier life-time limited current (Erhard et al. 2013; Prechtel et al. 2012a). All these currents are related to electric fields within a nanowire.

## 3.2. Photo-thermoelectric effect

The photo-thermoelectric effect is due to a local increase of the electron temperature induced by the laser illumination (Fu et al. 2011). A temperature difference $\Delta T$ at the interface of a nanowire and a second material, e.g. a metal contact, may result in a thermoelectric current:

$$I_{\text{Thermo}} = (S_{\text{nanowire}} - S_{\text{contact}})\Delta T/R,$$

with the Seebeck coefficients of the nanowire $S_{\text{nanowire}}$ and the metal contact $S_{\text{contact}}$ and $R$ the total resistance of the electrical circuit (Fu et al. 2011; Prechtel et al. 2012a; Varghese et al.). So far, the photo-thermoelectric effect has been observed in GaAs- (Prechtel et al. 2012a), InAs- (Erhard et al. 2013) and VO$_2$-based (Varghese et al.) nanowires with typical relaxation times below 10 ps (Erhard et al. 2013; Prechtel et al. 2012a). The contribution of the photo-thermoelectric current to the total photocurrent is often negligible, because the typical laser illumination intensities for photocurrent measurements raise the local temperature only by one tenths to a few Kelvin and typical Seebeck coefficients are of the order μV/K (Ahn et al. 2005; Fu et al. 2011). The photo-thermoelectric effect must not be confused with the bolometric effect (see section 3.3.4).

## 3.3. Photoconductance effects

As nanowires have a large surface-to-volume ratio, they are very sensitive to surface effects, such as the Fermi-level pinning leading to an accumulation or depletion layer at the nanowire surface (Hasegawa et al. 2002). The latter may induce a photoconductive gain (Ahmad et al.

2012; Ahn et al. 2007; Calarco et al. 2005; Chen et al. 2011; González-Posada et al. 2012; Hof et al. 2008; Kim et al. 2010; Polenta et al. 2008; Rossler et al. 2008; Sanford et al. 2010; Thunich et al. 2009; Zhai et al. 2010), a photodesorption effect (Calarco et al. 2011; den Hertog, M. I. et al. 2012; Harnack et al. 2003; Hsu et al. 2005; Huang et al. 2010; Hullavarad et al. 2009; Kind et al. 2002; Li et al. 2005; Pfüller et al. 2010; Prades et al. 2008; Soci et al. 2007; Suehiro et al. 2006) or even a persistent photoconductance (Ahmad et al. 2012; Calarco et al. 2005; Chen et al. 2011; González-Posada et al. 2012; Polenta et al. 2008; Rasool et al. 2012; Sanford et al. 2010; Winkelmann et al. 2007; Zhai et al. 2010). Furthermore, the temperature dependence of the conductivity may lead to a bolometric photoconductance effect (Tilke et al. 2003).

### 3.3.1. Photoconductive gain effect

As atoms at the crystal surface lack neighbors, surface reconstruction or adsorption of foreign atoms (see section 3.3.2) may occur. Hereby, new states at the surface are formed, which can lead to a Fermi-level pinning and a surface band bending (Ashcroft et al. 1987; Hasegawa et al. 2002; Spicer et al. 1979). The intrinsic electric field of the surface band bending points perpendicular to the nanowire axis. However, it can contribute two-fold to the photoresponse of nanowires: (i) for an up-bending (down-bending) at the surface, photogenerated holes (Rossler et al. 2008) (electrons) (Kim et al. 2010; Thunich et al. 2009) are trapped at the surface acting as a positive (negative) gate charge. In turn, the conducting part of the nanowire is increased (Thunich et al. 2009). This effect is sometimes called photogating effect. (ii) The photogenerated electrons (Ahmad et al. 2012; Ahn et al. 2007; Calarco et al. 2005; Rossler et al. 2008) (holes) (Kim et al. 2010; Thunich et al. 2009) increase the density of free charge carriers in the nanowire core which is called photodoping effect. The photogating effect was reported for GaAs- (Rossler et al. 2008; Thunich et al. 2009) and Ge-based (Kim et al. 2010) nanowires, the photodoping effect for GaAs- (Rossler et al. 2008; Thunich et al. 2009), Ge- (Ahn et al. 2007; Kim et al. 2010), GaN- (Calarco et al. 2005; Chen et al. 2011; González-Posada et al. 2012; Polenta et al. 2008; Sanford et al. 2010), InSe- (Zhai et al. 2010) and Si-based (Ahmad et al. 2012) nanowires. However, photogating and photodoping are interconnected to each other, because an interband excitation always generates electrons and holes in the nanowire.

### 3.3.2. Photodesorption effect

Photodesorption of chemisorbed or physisorbed foreign atoms can influence the conductivity of a nanowire as well (Figure 5). This photoconductance effect was observed in n-type semiconductor nanowires for chemisorbed oxygen (Harnack et al. 2003; Hsu et al. 2005; Hullavarad et al. 2009; Keem et al. 2004; Kind et al. 2002; Lee et al. 2004; Li et al. 2005; Prades et al. 2008; Suehiro et al. 2006; Yoon et al. 2010). In the dark, oxygen adsorbs at the nanowire surface by capturing free electrons: $O_2(g) + e^- \rightarrow O_2^-(ad)$ (Kind et al. 2002; Melnick 1957), which creates a bend bending upwards near the nanowire surface. In turn, a depletion layer is formed with low conductivity (Kind et al. 2002; Li et al. 2005; Soci et al. 2007). Exposing the nanowire to light may lead to photogenerated holes which migrate to the surface and discharge the adsorbed oxygen: $O_2^-(ad) + h^+ \rightarrow O_2(g)$ (Kind et al. 2002; Melnick 1957) (Figure 5a).

In literature two processes are reported. (i) For photons with an energy higher than the band gap of the nanowire, the remaining photogenerated electrons increase the conductivity of the nanowire core (Figure 5a) (Harnack et al. 2003; Hsu et al. 2005; Keem et al. 2004; Kind et al. 2002; Lee et al. 2004; Li et al. 2005; Soci et al. 2007; Yoon et al. 2010). The increased electron density may also reduce the width and height of Schottky barriers (Keem et al. 2004). (ii) For photons with an energy smaller than the band gap of the nanowire, electrons trapped at defect states related to the adsorbed oxygen are photoexcited to the conduction band. Hereby, the density of free charge carriers increases and the potential barriers are reduced (Figure 5b) (Keem et al. 2004; Li et al. 2005).

The photodesorption effect of oxygen has been shown for ZnO (Harnack et al. 2003; Hsu et al. 2005; Hullavarad et al. 2009; Kind et al. 2002; Li et al. 2005; Prades et al. 2008; Soci et al. 2007; Suehiro et al. 2006) , $SnO_2$ (Lee et al. 2004), CdSe (Yoon et al. 2010), ZnSe (Yoon et al. 2010), $WO_3$ (Huang et al. 2010) and GaN nanowires (Calarco et al. 2011; den Hertog, M. I. et al. 2012; Pfüller et al. 2010). In p-type ferroelectric semiconductor nanowires of SbSI, a negative photoconductive effect due to water desorption was found (Nowak et al. 2014).

### 3.3.3. Persistent photoconductance effects

Persistent photoconductivity describes a class of transport phenomena that prevail on a timescale of seconds to hours. There are several processes being classified as a persistent photoconductivity in literature.

(i) Surface band bending leads to a spatial separation of photogenerated holes and electrons (see Section 3.3.1) and in turn, to a reduced recombination rate (Ahmad et al. 2012; Calarco et al. 2005; Chen et al. 2007; Chen et al. 2011; González-Posada et al. 2012; Polenta et al. 2008; Sanford et al. 2010). This effect sensitively depends on the nanowire diameter (Calarco et al. 2005; Chen et al. 2011; González-Posada et al. 2012; Polenta et al. 2008) and it has been observed for GaN (Calarco et al. 2005; Chen et al. 2011; González-Posada et al. 2012; Polenta et al. 2008; Sanford et al. 2010), InSe (Zhai et al. 2010) and Si (Ahmad et al. 2012; Rasool et al. 2012; Winkelmann et al. 2007) nanowires.

(ii) Both, adsorption and desorption of molecules on a surface of the nanowire (see section 3.3.2) are rather slow processes. The increase and decrease of the photoconductivity can occur on a timescale of seconds to hours depending on ambient gas conditions and materials (Harnack et al. 2003; Hsu et al. 2005; Hullavarad et al. 2009; Keem et al. 2004; Kind et al. 2002; Nowak et al. 2014; Prades et al. 2008; Soci et al. 2007; Suehiro et al. 2006; Yoon et al. 2010).

(iii) Defect states in the nanowire core may also lead to a persistent photoconductivity. For ZnO nanowires, such a phenomena was explained via oxygen vacancy states which was excited to a metastable charged state (Hullavarad et al. 2009; Lany et al. 2005; Prades et al. 2008; Wang et al. 2011). An equivalent effect was reported for other II-VI and chalcopyrite Cu-III-VI2 semiconductors with anion vacancy states (Lany et al. 2005). In GaAs/AlGaAs nanowires with a Si delta-doping, deep DX centers can freeze out while capturing electrons at low temperatures (4.2 K) and the captured electrons can be activated via illumination leading again to a persistent photoconductance (Spirkoska et al. 2011).

### 3.3.4. Bolometric photoconductivity

The conductivity $\sigma$ of a semiconductor is a function of temperature (Sze et al. 2007). With absorption of photons, hot electrons are generated which redistribute their energy via electron-electron scattering processes on a femto- to picosecond timescale in the electron bath (Elsaesser et al. 1991; Neppl et al. 1979). This leads to an increased electron temperature $T_{el}$ relative to the lattice temperature $T_{lat}$ as the direct relaxation via phonon emission is less efficient (Neppl et al. 1979). There are three features which are characteristic for a bolometric photoconductivity: (i) The bolometric effect can be enhanced if the thermal coupling to the environment is reduced (Henini et al. 2002; Neppl et al. 1979). (ii) Typical relaxation times from 1 µs (Neppl et al. 1979) to 100 ms (Henini et al. 2002; Itkis 2006) are reported. (iii) The magnitude of the bolometric photoconductivity depends on the temperature derivative of the conductivity $d\sigma/dT$ (Henini et al. 2002; Itkis 2006). The bolometric effect was observed for a suspended Si nanowire in the Coulomb blockade regime (Tilke et al. 2003).

## 4. Absorption effects in semiconductor nanowires

By measuring the photocurrent, one can get information about absorption effects in single nanowires. Krogstrup et al. showed that the absorption of a single standing nanowire is enhanced due to its light concentrating property (cf. Figure 6) (Krogstrup et al. 2013) whereas a single Si nanowire with a diameter of 900 nm lying on a substrate can exhibit a partially coherent thin film absorption spectrum (Kelzenberg et al. 2008). Thinner Ge nanowires (10 nm – 110 nm) lying on a substrate have absorption peaks due to leaky mode resonances which are consistent with Lorenz-Mie theory for light scattering (Cao et al. 2009).

Because of the elongated geometry of nanowires, their absorption depends on the polarization of the exciting photons described by the polarization anisotropy $\rho = (I_\parallel - I_\perp)/(I_\parallel + I_\perp)$, with $I_\parallel$ ($I_\perp$) the photocurrent for parallel (perpendicular) polarized light to the nanowire. $\rho$ is generally positive for semiconductor nanowires as polarized light parallel to the nanowire can be better absorbed than perpendicularly polarized light. This can be explained by the dielectric mismatch of the vacuum and the nanowire material (Cao et al. 2009; Wang et al. 2001). When the wavelength $\lambda$ of the exciting light gets into the range of the nanowire diameter $d \approx \lambda/n_{\mathrm{NW}}$ the effect vanishes (Persano et al. 2011) ($n_{\mathrm{NW}}$ the refractive index of the nanowire). The polarization anisotropy can also influence the all-optical quantum interference control of electrical currents in single GaAs nanowires (Ruppert et al. 2010).

In addition, Fano resonances in Si nanostripe photodetectors have been investigated by measuring the scattering spectra (by bright- and dark field microscopy) and the absorption spectra (by photocurrent spectroscopy). Fan et al. found that the Fano resonance and by this the absorption resonance can be tuned by the nanostripe width (Fan et al. 2014).

Moreover, the recombination time of photogenerated charge carriers in a nanowire can be extracted from time-resolved photocurrent spectroscopy. In p-doped GaAs nanowires a recombination time of 1.5 ns was observed by THz time domain photocurrent spectroscopy (Prechtel et al. 2012a). This value is smaller than the 2.5 ns found by photoluminescence measurements (Demichel et al. 2010) as the recombination dynamics in photocurrent measurements is modified by the charge transport to the metal contacts (Prechtel et al. 2012a).

# 5. Morphologies explored by photocurrent spectroscopy

## 5.1. Crystal structures

Photocurrent spectroscopy can be used to investigate the band structure of a single semiconductor nanowire (Chen et al. 2013). Hereby, the band gaps of several types of nanowires have been characterized. InP zinc blende nanowires exhibit a band gap energy about 70 meV larger than for wurtzite nanowires at room temperature (Maharjan et al. 2009). P-doped GaAs nanowires have a room temperature band gap of about 1.42 eV (Thunich et al. 2009). The band gap of wurtzite $InAs_{1-x}P_x$ nanowires for different P contents has been measured by photocurrent spectroscopy, showing that wurtzite InAs (InP) has a band gap of 0.54 eV (1.65 eV) at 5 K (Trägårdh et al. 2007).

## 5.2. Surfaces and Interfaces

Photocurrents allow to probe local electric fields in the nanowires and thereby, surfaces and interfaces. For instance, surface band bending caused by Fermi-level pinning at the semiconductor surface (Sze et al. 2007) can lead to a Franz Keldysh effect (Franz 1958; Keldysh 1958) and by this to a sub-band gap absorption and photocurrent (cf. Figure 7). This effect is prominent for nanowires due to their large surface to volume ratio and it has been found for GaN- (Cavallini et al. 2007) and GaAs-based (Thunich et al. 2009) nanowires. Furthermore, photocurrent spectroscopy was used to investigate the homogeneity of nanowire surface states (Cavallini et al. 2006) and the associated persistent photocurrent in GaN- (Polenta et al. 2008) and ZnO-based (Hullavarad et al. 2009) nanowires.

By performing spectrally resolved scanning photocurrent microscopy, characteristics of Schottky contacts can be investigated. Schottky barriers can be extracted by following the Fowler model (Fowler 1931) using light below the band gap of the semiconductor. This prevents band to band excitations whereas excitations of charge carriers from the metal contact over the Schottky barrier can still occur (Sze et al. 2007). Yoon et al. showed that the Schottky barrier heights for phosphorous doped Si nanowires are reduced compared to bulk Si (Yoon et al. 2013). The advantage of the photocurrent method compared to current-voltage or capacity-voltage measurements is that no electric bias is applied which may change the population of interface states (Yoon et al. 2013).

The timescales of the underlying optoelectronic dynamics at nanowire surfaces and interfaces are too fast to be resolved by the conventional photocurrent spectroscopy. To this end, a time-resolved photocurrent spectroscopy can be applied (cf. subsection 2.2.2) as was performed for p-doped GaAs nanowires (Prechtel et al. 2012a) and for nominally undoped InAs nanowires (Erhard et al. 2013). Figure 8 shows the time-resolved photocurrent measurements on a single InAs nanowire performed with the measurement setup depicted in Figure 3. Each trace depicts the sampled, time-resolved photocurrent for scanning the pump laser along the nanowire from one metal contact (position 1) to the opposite contact (position 11) in eleven steps. At the nanowire metal interfaces, a thermoelectric current is observed (triangle) which decays in the first 10 ps after pulsed laser excitation (position 1 and 11). The dominant photocurrent contribution for long timescales is the drift of photogenerated holes (arrow in Figure 8). This drift current peak shifts in time for scanning the pump laser along the nanowire from one contact (position 1) to the second contact (position 11) (cf. Figure 3 and Figure 8a). From this time of flight analysis, one can deduce a hot hole drift velocity of about $3 \cdot 10^6$ cm/s which is consistent with the group velocity of the photogenerated holes (Erhard et al. 2013). The current of the photogenerated electrons can be detected in the first picoseconds where a THz oscillation due to a photo-Dember effect occurs (circle in Figure 8) which can be explained by the larger diffusivity of the photogenerated electrons compared to the one of the holes (Dember 1931).

GaAs nanowires can also exhibit a thermoelectric current if the nanowire-metal interfaces are nearly ohmic (Prechtel et al. 2012a). In the case of a Schottky contact, an ultrafast displacement current can occur with a full width of half maximum of 1.5 ps (Prechtel et al. 2012a).

## 5.3. Radial geometries

Recently, GaAs/AlGaAs core-shell nanowires have been increasingly investigated by photocurrent measurements as the AlGaAs shell reduces the Fermi-level pinning at the nanowire surface (Chang et al. 2012; Demichel et al. 2010; Parkinson et al. 2009; Rudolph et al. 2013). Additionally, the AlGaAs shell influences the polarization dependence since the material changes the dielectric confinement. The polarization anisotropy $\rho$ (cf. section 4) is larger for photon energies where only the GaAs core is excited compared to the case where the nanowire core and shell is excited (Persano et al. 2011). Kim et al. studied $\rho$ for excitation close to the GaAs core band gap. They found four different regions for $\rho$ in wurtzite GaAs/AlGaAs core-shell nanowires (Figure 9): (1) For excitation below the band gap of GaAs, photocurrents are dominated by defect states (Urbach tail) and the polarization anisotropy is positive due to dielectric mismatch (cf. section 4). (2) Near the band gap the Franz Keldysh effect becomes more pronounced and the polarization anisotropy is negative as it is dominated by the optical selection rules of the involved bands. (3) For higher photon energies, band to band transitions dominate the photocurrent and the dielectric mismatch leads to a positive polarization anisotropy (Kim et al. 2013). For low temperatures, the optical selection rules of the Franz Keldysh effect were used to determine the valence band splitting between heavy hole and light hole to be in the range of 90 meV (Kim et al. 2013). Combining photocurrent spectroscopy and photoluminescence measurements on GaAs/AlGaAs core shell nanowires, Chen et al. derived the band edge discontinuity in the conduction band $\Delta E_C$ between the GaAs core and the Al$_x$Ga$_{1-x}$As shell for different $x$, varying from about 0.17 meV for $x = 0.24$ to about 0.28 meV for a pure AlAs shell (Chen et al. 2013).

## 5.4. Axial geometries

Axial hetero-junction nanowires have the advantage that they can serve as a building block for multi-junction solar cells (Yao et al. 2014). Recently, Yao et al. reported an efficiency of 7.58% for GaAs-nanowire array solar cells by optimizing the array structure. A pin-junction close to the nanowire tip was observed to be favorable, as the carrier generation hot spot is located close to the nanowire tip due to the high absorption coefficient of GaAs. For higher photon energies, the surface recombination of the photogenerated charge carriers is enhanced resulting in a smaller external quantum efficiency for short wavelengths (Yao et al. 2014).

Moreover, zero dimensional quantum confined structures have been realized in axial geometries affecting the optical and the transport properties of a nanowire. Both can be investigated by photocurrent spectroscopy. This has been done for avalanche photodiodes consisting of an InP nanowire with a single quantum dot in the axial pn-junction (Figure 10). The photocurrent spectra in Figure 10c give evidence of the s, p and d-shell of the quantum dot, allowing to electrically detect a single exciton in a nanowire-based quantum dot (Bulgarini et al. 2012).

Rigutti et al. showed that the photocurrent to dark current ratio can be enhanced by introducing AlN quantum disks into a GaN nanowire. On the one hand, the introduced axial quantum disc heterostructure acts as a barrier for the dark current. On the other hand, piezoelectric fields support the extraction of photogenerated charge carriers from the quantum discs due to quantum confined stark effect (Rigutti et al. 2010).

## 5.5. Nanowire hybrid structures

For some applications, it can be beneficial to use a hybrid structure consisting of semiconductor nanowires and a second material. Over the last years, many hybrid structures have been introduced and investigated with regard to photovoltaic devices. The large interface in the nanowire hybrid structures is exploited for efficient separation of photogenerated electrons and holes. At the same time, the nanowires act as a good transport channels for photogenerated charge carriers to the electrodes (Leschkies et al. 2007; Ren et al. 2011; Shiu et al. 2010; Unalan et al. 2008). Leschkies et al. showed that the combination of ZnO nanowires with CdSe (Leschkies et al. 2007) or PdSe (Leschkies et al. 2009) quantum dots (nanoparticles) extends the device absorption and enhances the photon-to-current conversion efficiency. Hereby, the photogenerated electrons in the quantum dot are efficiently extracted and transported to the electrodes by the ZnO nanowires (Leschkies et al. 2007). A similar behavior was observed for nanowires combined with organic materials such as polymers (Ren et al. 2011; Shiu et al. 2010). Exploiting the fact that ZnO nanowires can be grown on single-walled carbon nanotube films, semiconductor nanowires are candidates for flexible photovoltaic devices (Unalan et al. 2008). Functionalizing nanowire surfaces with molecules may lead to new sensor applications. For example, a Si nanowire field effect transistor coated with porphyrin molecules enables the photoinduced electron transfer from the porphyrin to the nanowire as in artificial eyes (Winkelmann et al. 2007).  In addition, a nanowire can act as a near-field detector of optical plasmons. By performing polarization dependent photocurrent measurements, this was shown for an Ag nanowire crossing an electrically contacted Ge nanowire. Here, one uses the fact that a surface plasmon polariton in an Ag nanowire can generate electron hole pairs in a Ge nanowire (Falk et al. 2009).

## 6. Conclusion and Future trends

Semiconductor nanowires have unique absorption and photocurrent characteristics due to their geometry. Therefore, nanowires are good candidates for many future applications. The enhanced absorption in nanowires, the ability to build radial or axial pn-junctions, and the possibility to combine them with organic and nanoparticle solar cells may help to build cost-effective photovoltaic devices above the Shockley-Queisser limit. The capability to generate (Erhard et al. 2013; Seletskiy et al. 2011) and detect (Vitiello et al. 2012) THz radiation in nanowires makes them interesting for THz applications. Furthermore, nanowires are a possible building block for integrated photonic circuits as one can build LEDs or photodiodes based on a single nanowire (Tchernycheva et al. 2014). Nanowires exhibit a fast photoresponse down to the picosecond regime and they allow detecting single exciton states in quantum dots which can be exploited in future quantum technologies (Bulgarini et al. 2012). The small size and the principally small power consumption of nanowires makes it feasible to build devices which can be embedded in everyday life objects such as wrist watches or cell phones (Yang et al. 2010). To make this all come true, further improvement of the nanowire material and the nanowire device performance has to be achieved. Photocurrent measurements help to characterize and understand the nanowire devices and to identify future device geometries and applications.

# 7. Sources for further information

A summary regarding the optimization of nanowire photovoltaic devices can be found in (Garnett et al. 2011). Measurements with respect to photovoltaic performance on a single nanowire device are reviewed in (Tian et al. 2008). Further information about nanowire solar cells and thermoelectric devices can be found in (Hochbaum et al. 2010). Nanowire properties and devices with respect to photodetection applications are discussed in (Soci et al. 2010; VJ et al. 2011). Nanowire photodetectors and photovoltaic devices are reviewed together with other photonic nanowire devices in (Li et al. 2006; Yan et al. 2009). Further information about ultrafast carrier dynamics in semiconductor nanostructures can be found in (Shah 1996).

## Acknowledgements:

We thank L. Prechtel, S. Thunich, S. Hertenberger, P. Seifert, D. Spirkoska, H. Karl, C. Ruppert, M. Betz, G. Abstreiter, A. Fontcuberta i Morral, J. J. Finley, and G. Koblmüller for very fruitful collaborations and the ERC Grant NanoREAL (n°306754) for financial support.

# List of figure captions

## Figure 1
Characteristic timescales of photocurrent and photoconductance phenomena in semiconductor nanowires ranging from picoseconds to 1000s of seconds.

## Figure 2
Experimental setup to measure the photocurrent autocorrelation. A laser beam with laser pulses in the femtosecond regime is split into a pump and a probe pulse and they are subsequently recombined by a beam splitter (BS) (Erhard et al. 2013; Gabor et al. 2012; Urich et al. 2011). The time-delay between the pump and the probe pulse is adjusted by a translation stage in one of the two beam paths. Both laser beams are focused onto the electrically contacted nanowire, and the photocurrent is measured as a function of the time delay.

## Figure 3
Experimental setup for a THz time-domain photocurrent spectroscopy on a single semiconductor nanowire. Inset: optical microscope image of an InAs nanowire. Scale bar 4 µm. See text for further information. (Erhard et al. 2013) © 2012 by Wiley-VCH Verlag GmbH & Co. KGaA, Weinheim

## Figure 4
Nanowire geometries for homo- and hetero-junctions. a) Scanning electron microscope (SEM) image of a p-doped InP nanowire crossing an n-doped InP nanowire. Scale bar 2 µm. Reprinted by permission from Macmillan Publishers Ltd: Nature (Duan et al. 2001), copyright 2001. b) SEM image of a radial p(i)n-junction within a nanowire. Scale bar 100 nm. Reprinted by permission from Macmillan Publishers Ltd: Nature (Tian et al. 2007), copyright 2007. c) Schematic illustration of an axial p(i)n-junction nanowire. Reprinted with permission from (Kempa et al. 2008), copyright 2008 American Chemical Society. d) Schematic illustration of an nanowire-substrate junction.

## Figure 5
Schematic energy band diagram illustrating the photodesorption effect for illumination with photons having an energy a) higher and b) lower than the band gap energy. The band bending upwards at the nanowire surface is due to oxygen adsorption (negatively charged ions). Reprinted with permission from (Li et al. 2005). Copyright 2005, AIP Publishing LLC.

## Figure 6
Schematic illustration of the device geometry to measure the photocurrent of a single nanowire a) standing and b) lying on a $SiO_x$/Si substrate. c) The external quantum efficiency (EQE) for a lying nanowire (horizontal) and a standing nanowire (vertical). a) and c) Reprinted by permission from Macmillan Publishers Ltd: Nature Photonics (Krogstrup et al. 2013), copyright 2013. b) Reprinted by permission from Macmillan Publishers Ltd: Nature Materials (Cao et al. 2009), copyright 2009.

## Figure 7

a) SEM image of a p-doped GaAs nanowire bridging two gold electrodes. b) Corresponding spatially resolved photocurrent map ($V$ = +1 V, $T$ = 296 K, $\lambda$ = 780 nm, $I_{Laser}$ = 60 W/cm$^2$). c) Photocurrent $I_{Photo}$ as a function of excitation energy $E_{Photon}$ for Position I (circles) and II (squares) in a). Dashed line indicates the band gap of the GaAs nanowire ($V$ = -1 V, $T$ = 296 K, $I_{Laser}$ = 100 W/cm$^2$). Reprinted with permission from (Thunich et al. 2009). Copyright 2009, AIP Publishing LLC.

## Figure 8

Time-resolved photocurrent $I_{Sampling}$ for different pump laser positions along a single InAs-nanowire scanning the pump laser from one contact to the other (cf. inset Figure 3). Lines are fits to the data (Erhard et al. 2013). All data have an artificial offset for clarity. ($\lambda$ = 780 nm, $P_{Pump}$ = 6.4 kW/cm2, $P_{Probe}$ = 3.2 MW/cm2, $T_{Bath}$ = 77 K) (Erhard et al. 2013) © 2012 by Wiley-VCH Verlag GmbH & Co. KGaA, Weinheim.

## Figure 9

a) SEM image of a GaAs/AlGaAs core-shell nanowire before (top) and after the fabrication of gold contacts. Scale bar 1 µm. b) Photocurrent $I_{PC}$ and polarization anisotropy $\rho$ as a function of excitation energy $E$. Inset: Photocurrent $I_{PC}$ for a single wurtzite GaAs nanowire without shell. ($V$ = 5 V, room temperature, $P$ = 45 µW) Reprinted with permission from (Kim et al. 2013). Copyright 2013, AIP Publishing LLC.

## Figure 10

a) SEM image of the investigated InP nanowire device. Scale bar 1 µm. b) Schematics of carrier multiplication starting from an exciton generated in a nanowire QD, followed by tunneling in the nanowire avalanche region. c) Photocurrent spectroscopy at $V_{sd}$ = -2 V with 1 µW (InP data) and 20 µW (QD data) excitation powers. Band-edge absorption in the nanowire is observed around 825 nm. Absorption in the QD s-, p- and d-shells is observed at longer wavelengths. Reprinted by permission from Macmillan Publishers Ltd: Nature Photonics (Bulgarini et al. 2012), copyright 2012.

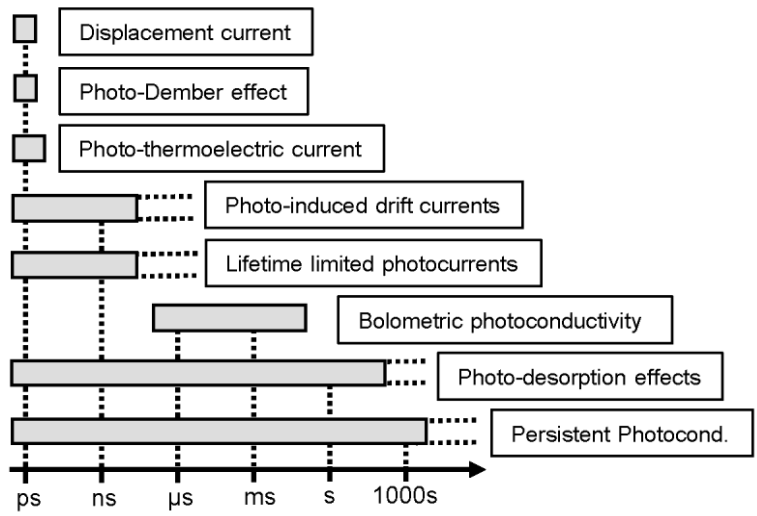

Figure 1

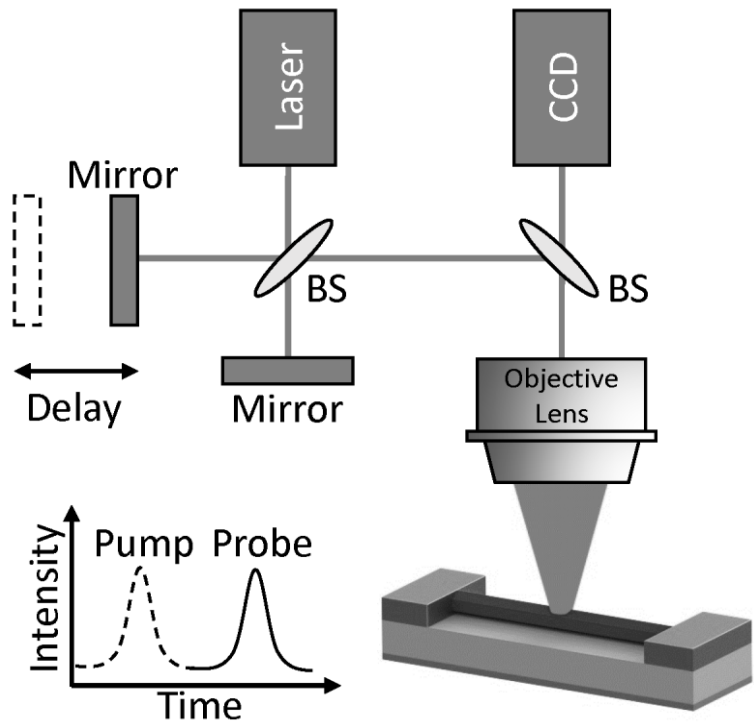

Figure 2

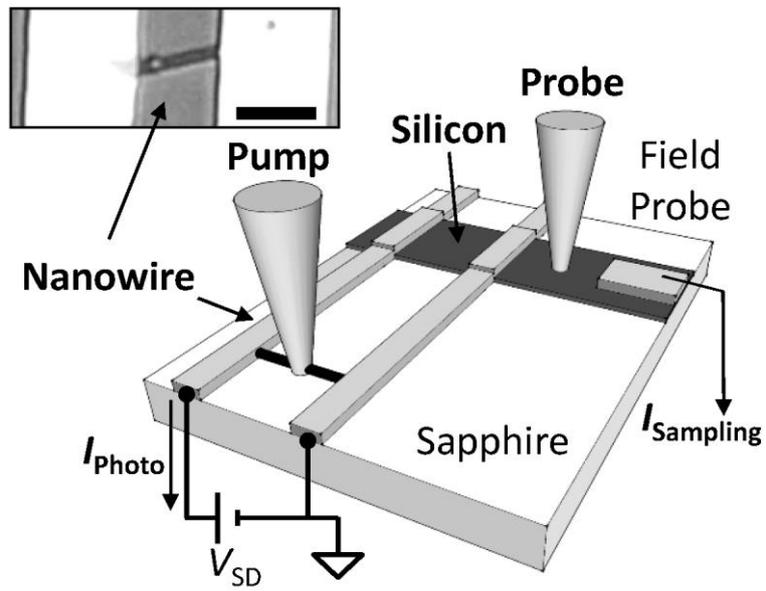

Figure 3

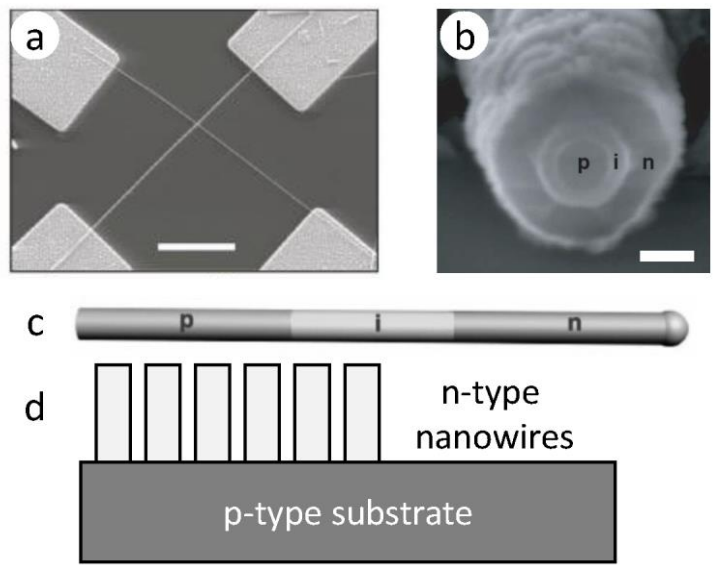

Figure 4

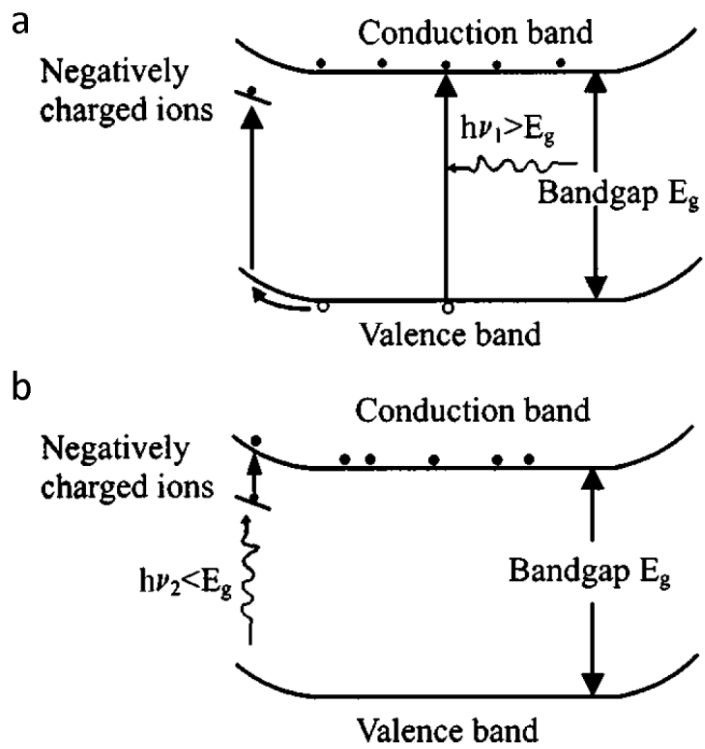

Figure 5

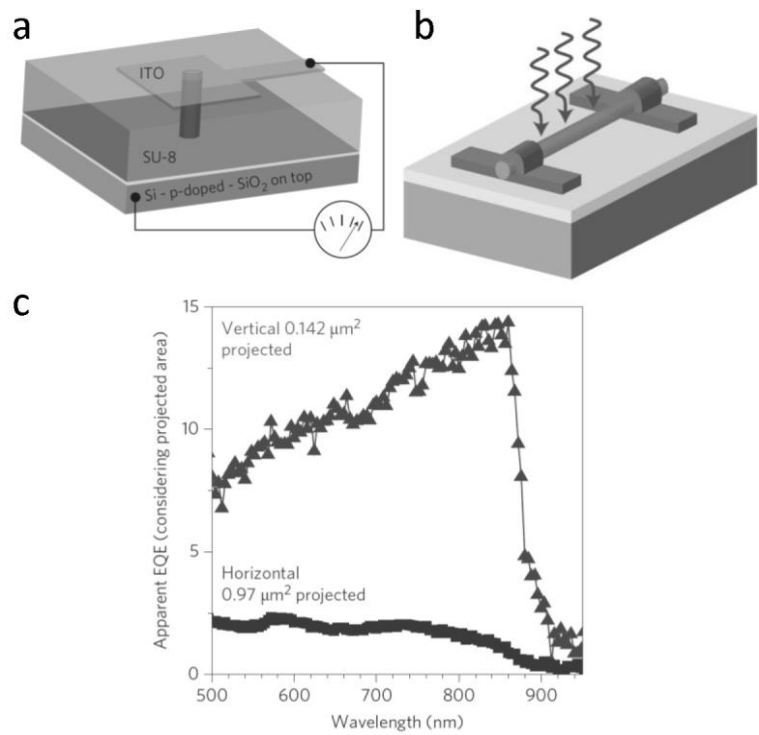

Figure 6

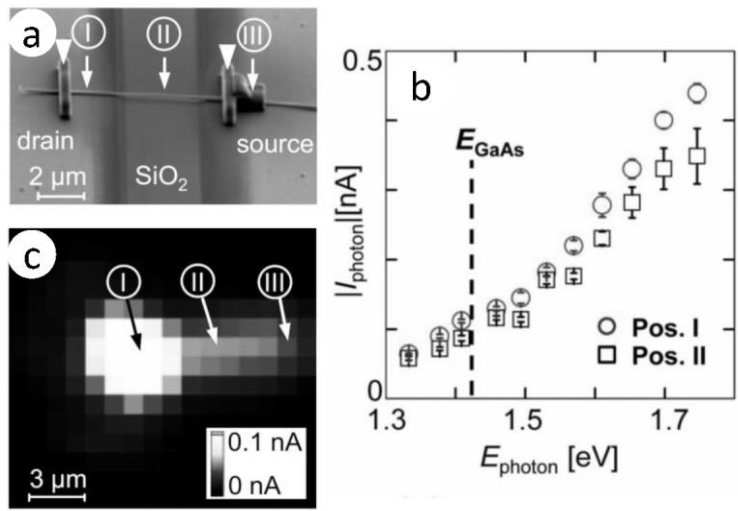

Figure 7

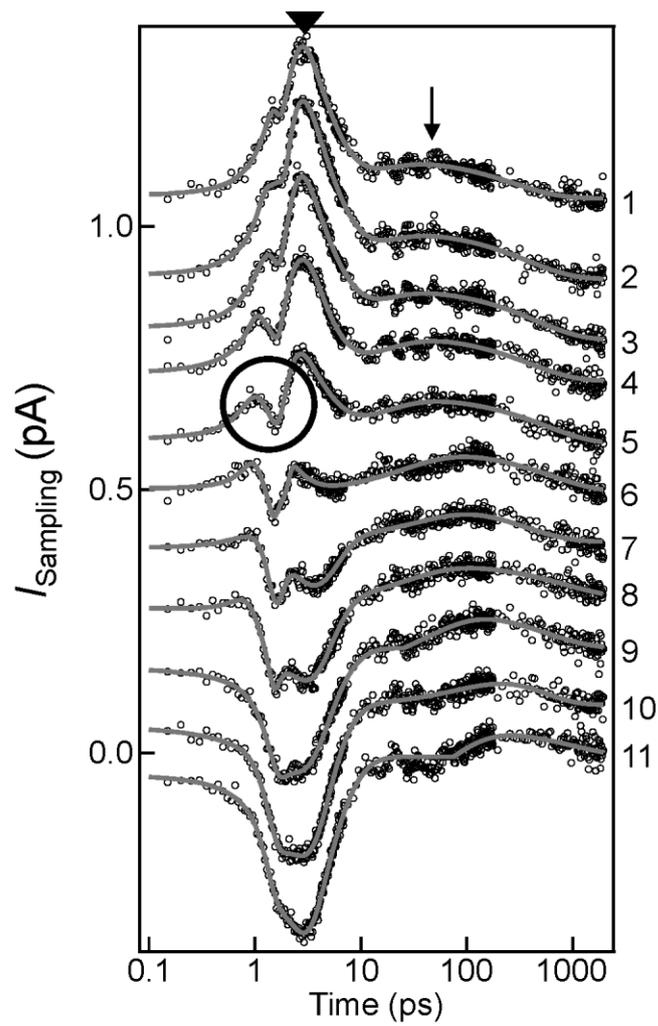

Figure 8

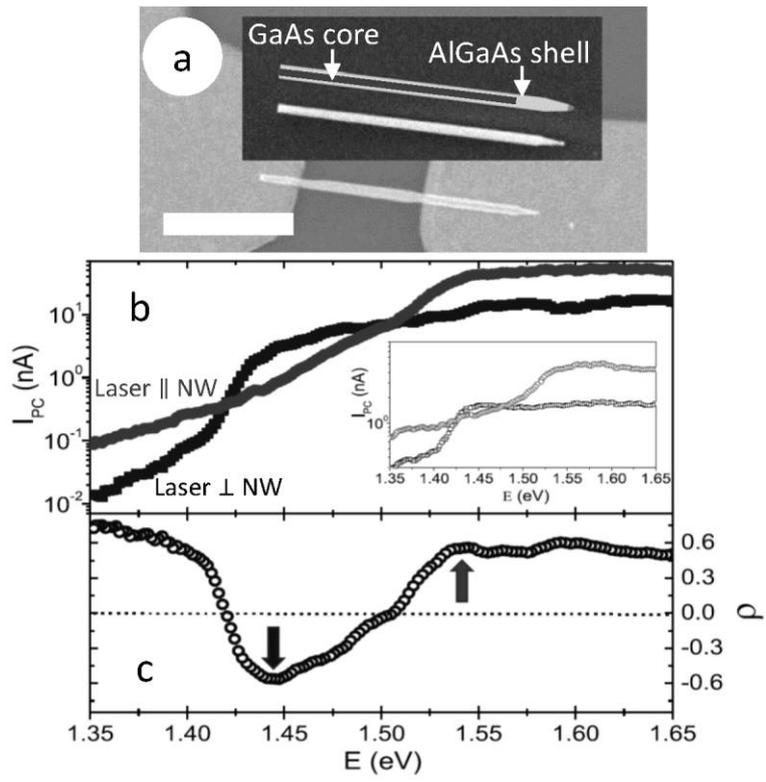

Figure 9

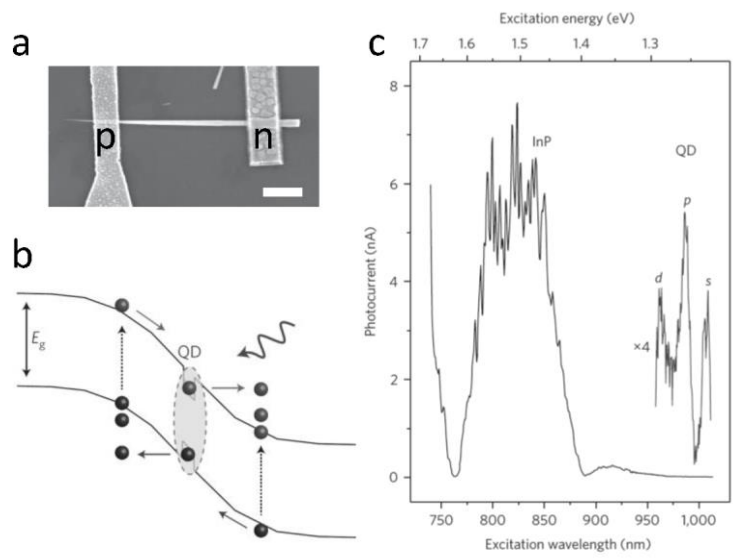

Figure 10